\documentclass[superscriptaddress,prd,preprintnumbers,twocolumn,nofootinbib]{revtex4}

\usepackage[utf8]{inputenc}
\usepackage{amsmath,amssymb,amsfonts}
\usepackage{bm, slashed, soul}
\usepackage{graphicx}
\usepackage{hyperref}
\usepackage{multirow}
\usepackage{comment}
\usepackage{color}
\definecolor{red}{rgb}{0.9, 0,0}

\newcommand{\lmk}{\left(}  
\newcommand{\rmk}{\right)}
\newcommand{\lkk}{\left[}  
\newcommand{\rkk}{\right]}

\newcommand{\del}{\partial}

\newcommand{\beq}{\begin{eqnarray}}
\newcommand{\eeq}{\end{eqnarray}}

\def\Mpl{M_{\rm pl}}

\newcommand{\be}{\begin{equation}}
\newcommand{\ee}{\end{equation}}

\newcommand{\bea}{\begin{array}}
\newcommand{\eea}{\end{array}}

\newcommand{\eq}[1]{Eq.~(\ref{#1})}

\def\beq#1\eeq{\begin{align}#1\end{align}}

\begin{document}

\preprint{
CTPU-PTC-20-07; \ 
TU-1100; \ 
MIT-CTP/5188 
}

\title{Spherically Symmetric Scalar Hair for Charged Black Holes}

\author{Jeong-Pyong Hong}
\affiliation{Center for Theoretical Physics of the Universe, Institute for Basic Science (IBS), 55 Expo-ro, Yuseong-gu, Daejeon 34126, Korea}
\affiliation{Center for Theoretical Physics, Department of Physics and Astronomy, Seoul National University, Seoul 08826, Korea}
\author{Motoo Suzuki}
\affiliation{Tsung-Dao Lee Institute, Shanghai Jiao Tong University, Shanghai 200240, China}
\author{Masaki Yamada}
\affiliation{Frontier Research Institute for Interdisciplinary Sciences, Tohoku University, Sendai, Miyagi 980-8578, Japan}
\affiliation{Department of Physics, Tohoku University, Sendai, Miyagi 980-8578, Japan}
\affiliation{Center for Theoretical Physics, Laboratory for Nuclear Science and Department of Physics, Massachusetts Institute of Technology, Cambridge, MA 02139, USA}

\date{\today}

\begin{abstract}
The no-hair theorem by Mayo and Bekenstein states that there exists no non-extremal static and spherical charged black hole endowed with hair in the form of a charged scalar field with a self-interaction potential. In our recent work~[\href{https://www.sciencedirect.com/science/article/pii/S0370269320301283}{\emph{ Phys. Lett. B } \textbf{803} 135324 (2020)}], we showed that the effect of a scalar mass term is important at an asymptotic infinity, which was omitted to prove the no-hair theorem. In this paper, we demonstrate that there actually exists static and spherical charged scalar hair, dubbed as Q-hair, around charged black holes, by taking into account the backreaction to the metric and gauge field. We also discuss that Q-cloud, which is constructed without the backreaction around a Reissner-Nordstr\"{o}m black hole, is a good approximation to Q-hair under a certain limit. 
\end{abstract}

\maketitle
\preprint{}


{\bf Introduction.--}
Direct observations of gravitational waves~\cite{Abbott:2016blz} and the black-hole shadow~\cite{Akiyama:2019cqa} have opened up a new era in black hole (BH) physics. 
As a unique candidate of the strong gravity regime, a better understanding of BH will be inevitable for a deeper understanding of gravity. Moreover, developments from both theoretical and experimental sides might provide a clue to a long-standing question about a unified description of general relativity and quantum mechanics.

Central to our understanding of BH nature relies on the no-hair theorem~\cite{Israel:1967wq,Carter:1971zc,Ruffini:1971bza}. In early stage~\cite{Bekenstein:1972ny,Bekenstein:1971hc,Bekenstein:1972ky,Bekenstein:1995un}, it precludes the static black hole with a scalar hair. Later on, however, black hole solutions with the scalar hair have been found, which includes the BHs with Skyrmion~\cite{Luckock:1986tr,Droz:1991cx}
Yang-Mills~\cite{Bizon:1990sr,Greene:1992fw,Maeda:1993ap,Mavromatos:1995kc}, axion~\cite{Campbell:1991rz,Mignemi:1992pm,Campbell:1990ai,Duncan:1992vz}, 
and Dilaton hairs~\cite{Kanti:1995vq}. For the recent review of the no-hair theorem, see Ref.~\cite{Herdeiro:2015waa}. 
The $tt$-component of the stress tensor is not equal to the $\theta\theta$-component of the stress tensor in these systems, which 
evades the no-hair theorem by Bekenstein~\cite{Bekenstein:1995un}. Although a system with a charged black hole has the same property, 
a stronger no-hair theorem, which excludes scalar hair around spherically symmetric charged black holes, is concluded by Mayo and Bekenstein~\cite{Mayo:1996mv}: 
\begin{quote}
{\it There exists no non-extremal static and spherical charged black hole endowed with hair in the form of a charged scalar field, whether minimally or nonminimally coupled to gravity, and with a regular positive semidefinite self-interaction potential.}
\end{quote}
In a recent work~\cite{Hong:2019mcj}, we reconsidered this no-hair theorem 
and argued that 
the above statement is not correctly concluded because they omitted a scalar mass term at an asymptotic infinity in the equation of motion of the scalar field. 

In this paper, we construct the numerical examples of spherical charged black hole with charged scalar hair, which are consistent with our argument. After the detailed explanation to our disproof, we demonstrate the numerical scalar hair solutions, dubbed as Q-hairs, in a polynomial scalar potential.%
\footnote{In  the flat spacetime, the attractive scalar self-interaction allows to form a non-topological soliton, known as Q-ball~\cite{Coleman:1985ki,Lee:1988ag,Kusenko:1997si,Kusenko:1997zq}.}
 Under certain limits, we also show that the scalar hairs are in accord with the cloud solutions~\cite{Hong:2019mcj} obtained under the Reissner Nordstrom BH background.

{\it Note added}: While completing our work, we became aware of an independent work by Carlos~A.~R.~Herdeiro and Eugen~Radu~\cite{Herdeiro:2020xmb}, who also demonstrate counter examples to the Mayo Bekenstein no-hair theorem.

\vspace{0.1cm}
{\bf Equations of motion.--}
We focus on a theory with a U(1) gauge field $A_\mu$ and a charged scalar field $\psi$ which minimally couples to the gravity. 
In this paper, we use the same notation as in Ref.~\cite{Mayo:1996mv} with $\xi = 0$ for simplicity. 
The action and the Lagrangian density are given by 
\beq
&S_{\rm SM}= \int \sqrt{-g}d^4 x \lmk - \frac{1}{16 \pi G} R + \mathcal{L}_M \rmk, 
\\
&\mathcal{L}_M=-\frac{1}{2}\left( (D^\alpha \psi)^* D_\alpha \psi +V(\psi,\psi^*)+\frac{1}{8\pi}F^{\alpha\beta}F_{\alpha\beta}\right), 
\eeq
where $D_\alpha=\partial_\alpha-iqA_\alpha$ and $F_{\alpha\beta}= A_{\beta, \alpha}-A_{\alpha,\beta}$
are the covariant derivative and the field strength of U(1) gauge interaction, respectively. 
The four-current density of electric charge is given by 
$j_\alpha = q {\rm Im} \lkk \psi^* D_\alpha \psi \rkk$ 
and the Maxwell equation is $F^{\alpha \beta}_{\ \  \ ; \beta} = 4 \pi j^\alpha$. 

We are interested in static solutions to the Einstein equation, 
$R^\alpha_\beta- R/2 \,\delta^{\alpha}_{\beta}=8\pi G\,T^\alpha_\beta$, 
with a charged black hole located at the center of the coordinate in an asymptotically flat spacetime. 
The metric is then written as 
\begin{align}
ds^2=-e^\nu dt^2+e^\lambda dr^2+r^2(d\theta^2+\sin^2\theta d\phi^2)\ ,
\end{align}
where $\nu$ and $\lambda$ are functions of $r$ 
and ${\cal O}(r^{-1})$ as $r \to \infty$. 
We define $r_{\rm H}$ by the radius of the event horizon at the surface of the BH, 
where $e^{- \lambda (r_{\rm H})} = 0$. 
We focus on the case of non-extremal black hole, in which 
$e^\nu, e^{-\lambda} = {\cal O}(r-r_{\rm H})$ for $r \to r_{\rm H}$.

Let us specify the gauge fixing of U(1) gauge symmetry. 
In a spherically-symmetric static spacetime, 
$F_{tr}$ is the only non-vanishing component for the field strength, 
which implies that only $A_t$ and $A_r$ are non-vanishing components. 
We can make $A_r = 0$ by a gauge transformation of $A_\alpha\to A_\alpha+\Lambda_{,\alpha}$ with $\Lambda=-\int A_r dr$. 
The time component of the gauge field $A_t$ must be the form of $f(r)+g(t)$ 
so that $F_{tr}$ is stationary. 
Then we can use a residual gauge transformation $\Lambda=-\int g(t)dt$ to make $A_t$ static. 
The scalar field must be in a form of $\psi=a(r)e^{ib(r)-i\omega t}$ with a real constant $\omega$ 
since otherwise the current and charge density depend on time. 
A further gauge transformation with $\Lambda=\omega t/q$ makes $\psi=a(r)e^{ib(r)}$ and $A_t\to A_t+\omega/q$.%
\footnote{
After the gauge transformation, 
$A_t$ in this paper corresponds to $g(r)/q$ ($\equiv A_0 + \omega / q$) in Ref.~\cite{Hong:2019mcj}. 
}
The conservation of charge implies that $b(r)$ is independent of $r$. 
Otherwise charge lead out continually to infinity. 
In summary, the non-vanishing components for the fields are $A_t (r)$ and $\psi = a(r)$.

The field equations and the Einstein equations are written as 
\begin{align}
\label{eq:1}
&a_{,rr}+\frac{1}{2}\left(\frac{4}{r}+\nu_{,r}-\lambda_{,r}\right)a_{,r}-(\dot{V}-q^2 e^{-\nu}A_t^2)e^{\lambda}a=0\ ,\\
\label{eq:2}
&A_{t,rr}+\frac{1}{2}\left(\frac{4}{r}-\nu_{,r}-\lambda_{,r}\right)A_{t,r}-4\pi q^2 a^2 e^\lambda A_t=0\ ,\\
\label{eq:3}
&e^{-\lambda}\left(\frac{1}{r^2}-\frac{\lambda_{,r}}{r}\right)-\frac{1}{r^2}=8\pi G T^t_t\ ,\\
\label{eq:4}
&e^{-\lambda}\left(\frac{\nu_{,r}}{r}+\frac{1}{r^2}\right)-\frac{1}{r^2}=8\pi G T^r_r\ ,
\end{align}
where $(t,t)$ and $(r,r)$ components of the energy-momentum tensor are given by 
\begin{align}
\label{eq:t00}
8 \pi T^t_t= 4 \pi \lmk - e^{-\lambda} a^{2}_{,r}- e^{-\nu} q^2A_t^2\,a^2 - V \rmk - e^{-\nu-\lambda}A_{t,r}^{2},\\
\label{eq:t11}
8 \pi T^r_r= 4 \pi \lmk e^{-\lambda} a^{2}_{,r}+ e^{-\nu} q^2A_t^2\,a^2 - V \rmk - e^{-\nu - \lambda} A_{t,r}^{2}, 
\end{align}
where $\dot{V} \equiv \del V / \del a^2$. 

We define $Q (r)$ by the electric charge enclosed by the sphere of radius $r$ 
such as $Q(r) \equiv Q_{\rm BH} + Q_\psi (r)$, 
where $Q_{\rm BH}$ is the charge of the BH 
and 
\beq
 Q_\psi (r)
 = 
 - 4 \pi q^2 \int_{r_{\rm H}}^r dr' \, r'^2 e^{(\lambda-\nu)/2} 
 a^2 (r) A_t (r), 
 \label{Q_psi}
\eeq
is the electric charge of $\psi$ enclosed by the sphere of radius $r$. 
Then we find 
\beq
 e^{-(\nu+\lambda)/2} A_{t,r} = - \frac{Q(r)}{r^2}, 
 \label{Q1}
\eeq
from the Gauss's law.

\vspace{0.1cm}
{\bf No hair theorem by Mayo and Bekenstein.--}
Before we are going to dispute the no-hair theorem, let us briefly review the proof by Mayo and Bekenstein~\cite{Mayo:1996mv}. 
Noting that 
$e^{-(\nu+ \lambda)/2} \approx ({\rm const.})$ 
near the event horizon, 
Eq.~(\ref{Q1}) implies that $A_{t,r}$ is regular at $r = r_{\rm H}$. 
The asymptotic form of the fields and the metric near the event horizon $r = r_{\rm H}$ 
are then written as 
\begin{align}
\label{eq:kpat}
&A_t=c_0-c_1(r-r_{\rm H})\ ,\\
&e^\nu=c_2(r-r_{\rm H})\ , \quad 
e^\lambda=\frac{c_3}{r-r_{\rm H}}\ ,
\label{eq:kpB}
\end{align}
where $c_i$ denote positive finite constants. 
Note that $c_1$ is determined by \eq{Q1}.

Now, suppose that the gauge field does not vanish at an asymptotic infinity, $A_t(\infty)\neq0$. 
Then $a$ must vanish asymptotically to satisfy the Maxwell equation \eq{eq:2}. 
Since we consider an asymptotically flat spacetime, 
we require $V(a) \to 0$ and $a \to 0$ for $r \to \infty$. 
Then one may think that \eq{eq:1} reduces to 
\begin{align}
a_{,rr}+\frac{2}{r}a_{,r}+q^2A_t(\infty)^2a=0  
\ \ {\rm for} \ r \to \infty. 
\label{eq:asymptotic}
\end{align}
As we show in the next section, this is correct only if we omit the scalar mass term. 
For a moment, we assume the above equation, following Ref.~\cite{Mayo:1996mv}. 
The solution to the equation has the form of 
\begin{align}
a\sim\frac{1}{r}\sin\left(qA_t(\infty)r+\chi\right) 
\ \ {\rm for} \ r \to \infty, 
\end{align}
where $\chi$ is a constant. 
In this case, the electric charge density of the scalar field, $\sqrt{g_{tt}} j^t$, is given by 
\begin{align}
 \sqrt{g_{tt}} j^t (r) \sim 
 - 4 \pi q^2 
   A_t (\infty) \sin^2 \left(qA_t(\infty)r+\chi\right) 
 \ \ {\rm for} \ r \to \infty, 
\end{align}
Then the total electric charge diverges, 
which 
means that 
the assumption of $A_t(\infty)\neq 0$ does not lead to a physical solution. 
One thus concludes that 
\begin{align}
A_t(\infty)=0\ . 
\label{Atinf}
\end{align}

Once $A_t (\infty) = 0$ is obtained, 
we can show that $A_t(r)$ is a monotonic function. 
Suppose that $A_t (r)$ has an extremal at a certain radius $r_*$. 
Then \eq{eq:2} implies that $A_t'' (r_*)$ and $A_t (r_*)$ has the same sign 
and hence the extremal is a minimum for $A_t (r_*) > 0$ and is a maximum for $A_t (r_*) < 0$. 
This can not be consistent with $A_t (\infty) = 0$, so that we conclude that $A_t(r)$ is a monotonic function. 
Since the overall sign of $A_t$ can be changed by changing the sign of $q$ without loss of generality, 
we can set $A_t (r) > 0$. In this notation, $A_t(r)$ is a monotonically decreasing function. 
In particular, $A_t (r_{\rm H})$ ($\equiv c_0$) must be nonzero and positive.

Given these properties, 
the second term in \eq{eq:2} is finite for $r\to r_{\rm H}$ from Eq.\,\eqref{eq:kpat}-Eq.\,\eqref{eq:kpB}. 
In the third term, $A_t (r_{\rm H})$ is finite and 
$e^\lambda$ diverges as $1/(r-r_{\rm H})$ from Eq.\,\eqref{eq:kpB}, 
so that $a$ must behave as $a= 0$ for $r=r_{\rm H}$.
On the other hand, 
Eq.\,\eqref{eq:1} can be approximated to be 
\begin{align}
\label{eq:scalarhorizon1}
a_{,rr}+\frac{1}{r-r_{\rm H}}a_{,r}+\frac{q^2 c_0^2 c_3}{c_2 (r-r_{\rm H})^2}a=0\ \ {\rm for} \ r \to r_{\rm H}. 
\end{align}
The solution to this equation is given by 
\begin{align}
a=B\sin \lkk q c_0 c_3^{1/2} c_2^{-1/2} {\rm ln} \lmk \frac{r-r_{\rm H}}{D} \rmk  \rkk, 
\end{align}
where $B$ and $D$ are arbitrary constants. This solution is, however, inconsistent with the condition of $a = 0$ for $r = r_{\rm H}$ 
because there is no choice of constants to satisfy $a\to 0$ for $r\to r_{\rm H}$. 
This means that there is no solution for the equations of motion 
and one may thus conclude the no-hair theorem for a spherically-symmetric static black hole. 
However, as we briefly noted below \eq{eq:asymptotic}, 
the above argument is correct only if we omit the scalar mass term in \eq{eq:asymptotic}.

\vspace{0.1cm}
{\bf Incompleteness of the no hair theorem.--}
Now we show that the no-hair theorem is not applicable to 
the case in which the complex scalar field has non-zero mass~\cite{Hong:2019mcj}. In the next section, 
we explicitly show our numerical solutions of scalar hair for a polynomial potential.

In the above proof, we cannot deduce the condition of Eq.\,\eqref{eq:kpat} if the mass of the scalar field is non-negligible. 
Even if $V(\psi)$ and $\psi$ are asymptotic to $0$ for $r \to \infty$, 
we must include the mass term because it is in the same order with the last term in \eq{eq:asymptotic}. 
Indeed, the asymptotic scalar equation in Eq.\,\eqref{eq:asymptotic} is modified to
\begin{align}
a_{,rr}+\frac{2}{r}a_{,r}-(\mu^2-q^2 A_t(\infty)^2) a=0 
\ \ {\rm for} \ r \to \infty, 
\end{align}
where $\mu^2$ is the mass squared for the scalar field. 
If the parenthesis is positive, 
the solution is given by 
\beq
 a \sim \frac{1}{r} e^{- \sqrt{\mu^2- q^2 A_t^2 (\infty)} \, r}, 
\eeq
and the total electric charge is finite. 
Therefore, there may be a consistent solution even for $A_t (\infty) \ne 0$ 
and \eq{Atinf} is not necessarily true. 
In particular, 
there is no reason that we cannot take $A_t(r_{\rm H}) = 0$ 
and $a(r_{\rm H}) \ne 0$.

In fact, there is a consistent solution if we take
\begin{align}
A_t = {\cal O} (r-r_{\rm H})~~\text{for $r\to r_{\rm H}$}\ .
\label{BCAt}
\end{align}
Then the third term in \eq{eq:2} can be finite even if $a$ is finite for $r \to r_{\rm H}$. 
From Eq.\,\eqref{eq:1}, 
one can check that $a_{,r}$ at $r = r_{\rm H}$ is finite and is given by 
\begin{align}
a_{,r} (r_{\rm H}) =\frac{\dot{V}a}{r_{\rm H}\left(\frac{1}{r_{\rm H}^2}+4\pi G\left(-V-\frac{Q_{\rm BH}^2}{4\pi r_{\rm H}^4}\right)\right)}\ .
\label{BCap}
\end{align}
Here we used $e^{-\nu} A_t^2 \to 0$ for $r \to r_{\rm H}$ 
and \eq{Q1} with $Q(r_{\rm H}) = Q_{\rm BH}$.

Note that $A_t \to 0$ for $r \to r_{\rm H}$ 
is the condition that is used to find a static solution of Q-cloud in Ref.~\cite{Hong:2019mcj}. 
It is also known to be at the threshold for superradiance~\cite{Bekenstein:1973mi, Herdeiro:2014goa}, 
which is also the case for scalar hairs around Kerr BH~\cite{Hod:2012px, Hod:2012zza, Herdeiro:2014goa, Herdeiro:2014pka, Hod:2014baa, Benone:2014ssa, Herdeiro:2015tia, Huang:2016qnk, Huang:2017whw, Herdeiro:2017oyt, Huang:2019xbu}. 
In the next part, we search numerical solutions for the scalar hairs 
that have the above asymptotic forms.

\vspace{0.1cm}
{\bf Numerical solutions.--}
Since the effect of scalar hair is expected to be negligible near the surface of the event horizon, 
we expect that the metric near the BH surface 
should be written in the form of the Reissner-Nordstrom BH with a nonzero vacuum energy such as 
\beq
 e^{-\lambda} \approx 1 - \frac{2G M_{\rm BH}}{r} + \frac{G Q_{\rm BH}^2}{r^2} - \frac{8 \pi G \Lambda r^2}{3}, 
 \qquad {\rm for} \   r \approx r_{\rm H}, 
 \label{SdS}
\eeq
where $\Lambda = V(\psi (r_{\rm H})) / 2$. 
One can check that $\lambda_{,r} (r_{\rm H})$ derived from \eq{SdS} is consistent with 
the one calculated from \eq{eq:3}. 
Since $e^{- \lambda} = 0$ for $r = r_{\rm H}$, 
we then obtain 
\beq
 M_{\rm BH} = \frac{r_{\rm H}}{2G} + \frac{Q_{\rm BH}^2}{2 r_{\rm H}} - \frac{4 \pi r_{\rm H}^3}{3} \Lambda. 
 \label{MBH}
\eeq
If $r_{\rm H}$ is given, \eq{MBH} can be regarded as the definition of the BH mass $M_{\rm BH}$. 
Conversely, one may specify $M_{\rm BH}$ and determine $r_{\rm H}$ from \eq{MBH}.

For the purpose of numerical simulation, 
it is convenient to define $E(r)$ by 
\beq
 e^{-\lambda} \equiv 1 - \frac{2 G E(r)}{r}. 
 \label{Er}
\eeq
Then it satisfies 
\beq
 \del_r E(r) = - 4 \pi r^2 T^t_t, \quad 
 E(r_{\rm H}) = \frac{r_{\rm H}}{2G}. 
\eeq
We note that a boundary condition of $e^{-\lambda} \to 1$ for $r \to \infty$ 
is manifestly satisfied in \eq{Er}. 
The function $E(r)$ is just 
the total energy enclosed by the sphere of radius $r$: 
\beq
 &E(r) = M_{\rm BH} + E_A + E_\psi, 
 \\
 &E_A = - \frac{Q_{\rm BH}^2}{2 r_{\rm H}} + \frac12 \int_{r_{\rm H}}^r d r \, r^2 \frac{Q^2(r)}{r^4}, 
 \label{E_A}
 \\
 &E_\psi = \frac{4 \pi r_{\rm H}^3 \Lambda}{3}  + 4 \pi \int_{r_{\rm H}}^r d r \, r^2 \frac12 \lmk 
  e^{-\lambda} a^{2}_{,r}+ e^{-\nu} q^2A_t^2\,a^2 + V
 \rmk. 
 \label{E_psi}
\eeq

\begin{figure}
\centering
  \includegraphics[width=0.9\linewidth]{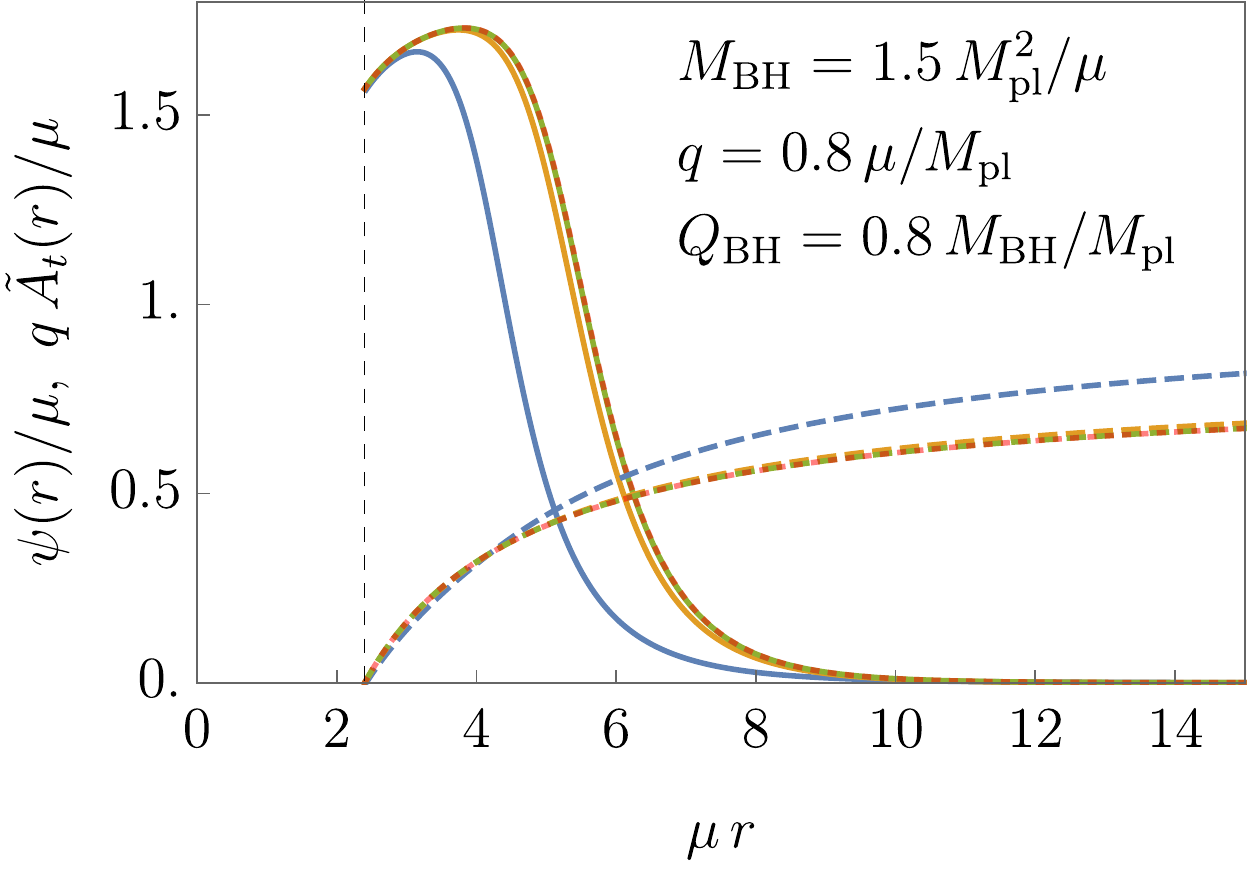}
   \caption{Solutions of $\psi(r)$ (solid lines) and $q \tilde{A}_t (r)$ (dashed lines) in the unit of $\mu$. 
   We take $M_{\rm BH} = 1.5 \Mpl^2 / \mu$, $q = 0.8 \mu / \Mpl$, and $Q_{\rm BH} = 0.8 M_{\rm BH} / \Mpl$ with $\mu / \Mpl = 0.02$ (blue), $0.005$ (orange), and $0.001$ (green). 
   We also show the solutions for the case without the backreaction to the metric and gauge field 
   as the red dotted lines, which are completely overlapped with the case of $\mu / \Mpl = 0.001$. }
\label{fig1}
\end{figure}

We would also like to impose the boundary condition of $\nu$ 
so that the spacetime is a Minkowski spacetime at the asymptotic infinity, 
namely, $\nu(r) \to 0$ as $r \to \infty$. 
For the purpose of numerical simulation, 
we solve the above equations for an arbitrary $\nu(r_{\rm H})$ 
and determine $\nu(\infty)$, which is nonzero in general. 
Then we rescale the time variable by $\tilde{t} = e^{ \nu (\infty)/2} t$ 
so that a new $\tilde{\nu}$ ($\equiv \nu (r) - \nu(\infty)$) is asymptotic to $0$ as $r \to \infty$. 
We should also rescale 
$\tilde{A_t} = e^{- \nu (\infty)/2} A_t$ 
with $q$ fixed, so that $q \tilde{A_t}(\infty)$ can be regarded as a chemical potential of the Q-hair. 

Using the shooting method, we ensure 
the field value $\psi(r)$ to be asymptotic to $0$ for $r \to \infty$. 
Free parameters are $q$, $Q_{\rm BH}$, $r_{\rm H}$ (or $M_{\rm BH}$), 
and parameters in the scalar potential, such as the scalar mass $\mu$. 
Following the previous section, we explore scalar hair solutions assuming boundary conditions at the horizon, Eqs.~(\ref{BCAt}) and (\ref{BCap}). 
The scalar potential is taken to be the following polynomial form: 
\begin{align}
V=\mu^2a^2- \frac{\lambda}{2} \,a^4+ \frac{\alpha}{4}\,a^6\ ,
\end{align} 
where 
$\mu$ is the scalar mass at the potential minimum, 
$\lambda$ and $\alpha$ are couplings. 
We take $\lambda = 1$ and $\alpha = \lambda^2/ (3 \mu^2)$ as an example.

\begin{figure}[tttttt]
\centering
  \includegraphics[width=0.9\linewidth]{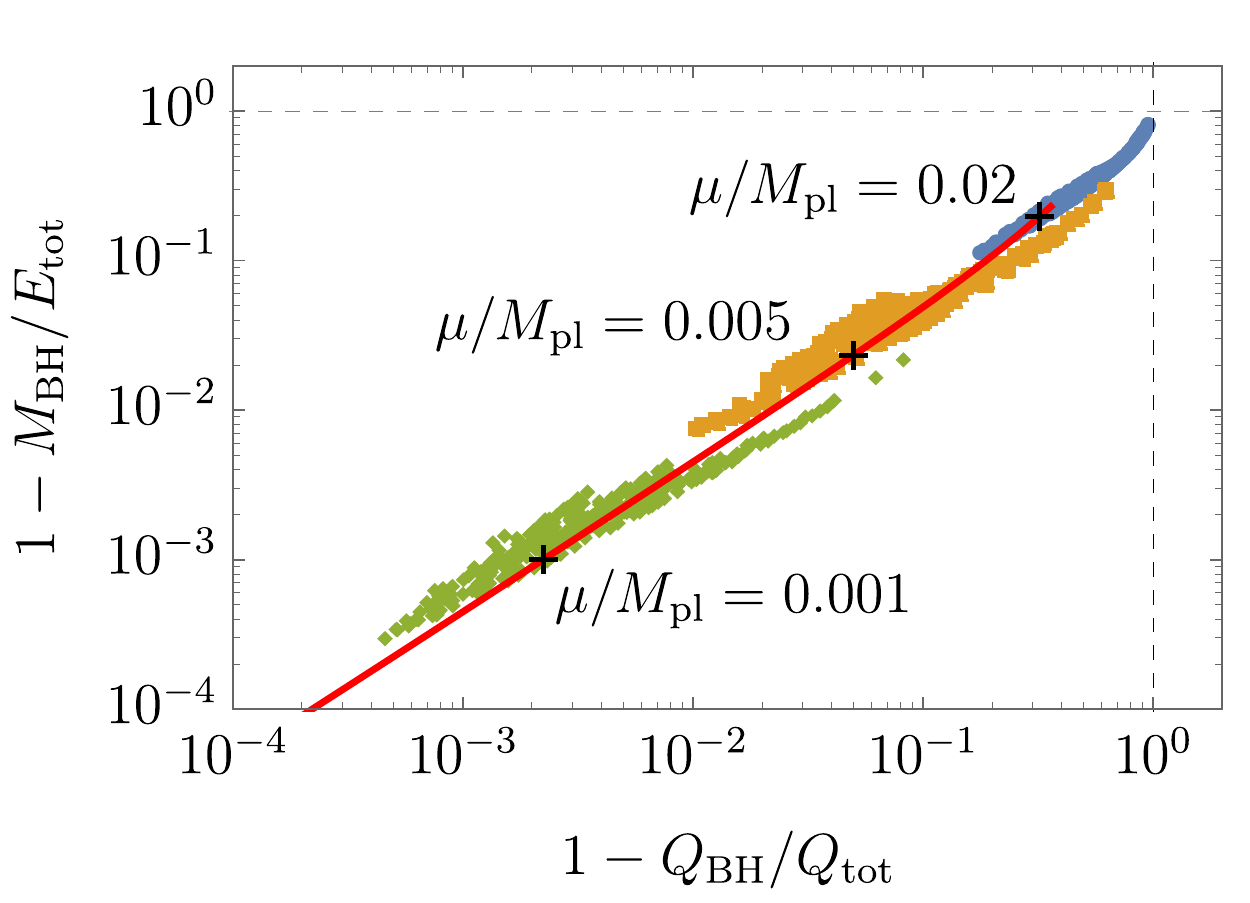}
   \caption{$(1 - M_{\rm BH} / E_{\rm tot})$-$(1-  Q_{\rm BH} / Q_{\rm tot})$ plot for the existence of Q-hair. Each dot represents the parameter at which we find a Q-hair solution. 
   We take $\mu / \Mpl = 0.02$ (blue), $0.005$ (orange), and $0.001$ (green). 
   The red line represents one-parameter family of solutions 
   for $\mu/\Mpl$ ($\lesssim 0.023$ with $M_{\rm BH} = 1.5 \Mpl^2 /\mu$, $q = 0.8 \mu / \Mpl$, 
   and $Q_{\rm BH} = 0.8 M_{\rm BH} / \Mpl$. 
   The black crosses represent the three examples that are plotted in Fig.~\ref{fig1}. 
   }
\label{fig2}
\end{figure}

First, suppose that $(1 - M_{\rm BH} / E_{\rm tot}) \ll 1$ and $(1-  Q_{\rm BH} / Q_{\rm tot}) \ll 1$, 
namely 
$Q_{\rm BH} \gg Q_\psi (\infty)$ 
and $M_{\rm BH} \gg E_A (\infty) + E_\psi (\infty)$, 
where $E_{\rm tot} \equiv E(\infty)$ and $Q_{\rm tot} \equiv Q(\infty)$. 
In this case 
we can neglect the right-hand side of Eqs.~(\ref{eq:3}) and (\ref{eq:4}) 
and the solution to those equations are just given by 
the one for the Reissner-Nordstr\"{o}m BH: 
$\nu = - \lambda = \ln (r^2 - 2 M_{\rm BH} r + Q_{\rm BH}^2) / r^2)$. 
If the third term of \eq{eq:2} is negligible, 
the solution of the gauge field is then given by $A_t(r) = Q_{\rm BH} (1/r - 1/r_{\rm H})$, 
where we assume $A_t(r_{\rm H}) = 0$. 
The only non-trivial equation is \eq{eq:1}, 
which can be numerically solved by the shooting method. 
This has been done in Ref.~\cite{Hong:2019mcj} 
and there actually exist solutions of scalar hair. 
We note that 
these limits can be realized by $\mu /\Mpl \to 0$ 
with $c_Q$, $c_M$, $c_q$ fixed, 
where $G \equiv 1/ \Mpl^2$ and 
\begin{align}
\label{eq:rescale}
Q_{\rm BH} \equiv c_Q \frac{M_{\rm BH}}{\Mpl},~ M_{\rm BH} \equiv c_M \frac{\Mpl^2}{\mu},~ 
q \equiv c_q \frac{\mu}{\Mpl}. 
\end{align}
As we have confirmed that there are solutions to \eq{eq:1} in this limit, 
we can start from a small $\mu/\Mpl$ and increase it to find numerical solutions of Q-hair 
in the full equations.

We show three examples of Q-hair in Fig.~\ref{fig1}, 
where we take $c_M = 1.5$ and $c_Q = c_q = 0.8$. 
The solid lines represent $\psi(r) / \mu$ 
while the dashed lines represent $q \tilde{A}_t (r) / \mu$ 
for the case of $\mu / \Mpl = 0.02$ (blue), $0.005$ (orange), and $0.001$ (green). 
We also show the Q-cloud solution around a Reissner-Nordstr\"{o}m BH with the same parameters 
as red dotted lines. 
We cannot distinguish between 
the green lines and the red dotted lines 
as they are completely overlapped. 
This means that the Q-cloud solution is a good approximation 
for $\mu / \Mpl \lesssim 0.001$ in this case. 

In Fig.~\ref{fig2}, 
each dot represents a parameter 
at which there exists Q-hair solution. 
We fix $\mu/ \Mpl$ as $0.02$ (blue dots), $0.005$ (orange squares), 
and $0.001$ (green diamonds) 
with $c_q = 0.8$. 
We take $c_M$ and $c_Q$ randomly within $(0, 10)$ and $(0,1)$, respectively, 
and no solution is found for $c_M \gtrsim 9$. 
We can see that 
Q-hairs with smaller $\mu / \Mpl$ 
have smaller $(1 - M_{\rm BH} / E_{\rm tot})$ and $(1-  Q_{\rm BH} / Q_{\rm tot})$. 
This implies that the backreaction to the metric is negligible for a small $\mu / \Mpl$. 

The red line in Fig.~\ref{fig2} 
is a one-parameter solution for $\mu / \Mpl$ 
with $c_M = 1.5$, $c_Q = 0.8$, and $c_q = 0.8$. 
It is bounded above as no solution is found for $\mu / \Mpl \gtrsim 0.023$ in this case. 
The three black crosses represent the parameters 
corresponding to the three examples used in Fig.~\ref{fig1}, 
namely for $\mu / \Mpl = 0.02$, $0.005$, and $0.001$. 
As we expect from Eqs.~(\ref{Q_psi}), (\ref{E_A}), and (\ref{E_psi}), 
both $(1 - M_{\rm BH} / E_{\rm tot})$ and $(1-  Q_{\rm BH} / Q_{\rm tot})$ 
are almost proportional to $(\mu / \Mpl)^2$ 
for a small $\mu / \Mpl$. 
We conclude that the Q-cloud solutions 
are good approximation in the limit of $\mu / \Mpl \to 0$. 
In particular, this implies that the vacuum energy from the scalar potential is not necessary for the existence of hairy solution around a charged BH. 
This is in contrast to the axion-hair reported in Ref.~\cite{Mavromatos:2018drr}. 
In our case, the repulsive force that is balanced by the gravitational attractive force 
mainly comes from the U(1) gauge interaction. This is the reason that the vacuum energy, that provides another repulsive force, does not play an important role in our system.

We note that our Q-hair is secondary in the sense that it does not introduce new physical parameters in the solutions~\cite{Coleman:1991jf}. The hair is not an independent quantum number from the mass and charge.

\vspace{0.1cm}
{\bf Discussion.--} 
We have seen that some scalar hair solutions match well with the solutions obtained without including the backreaction to the metric and gauge field, particularly in the limit of $\Mpl \to \infty$. 
This is phenomenologically important because, e.g., a typical grand-unified scale is three orders of magnitude smaller than the Planck scale, which is small enough for the backreaction to the metric is negligible.

Although the stability of Q-cloud against small perturbations as well as non-perturbative process is justified in Ref.~\cite{Hong:2019mcj}, 
that of Q-hair has not been explored yet. We need to investigate 
the behavior of small perturbations on top of the Q-hair to ensure its stability. 
This would be an interesting direction to future work. 
We note that the instability analysis of Yang-Mills hair around a BH reported in Ref.~\cite{Mavromatos:1995kc} is not directly applicable to our case because they consider a wine-bottle potential for a scalar field, which does not respect a Q-ball solution even in a flat spacetime.


\begin{acknowledgments}
J.~P.~H is supported by Korea NRF-2015R1A4A1042542 
and IBS under the project code, IBS-R018-D1. 
M.~Y. was supported by JSPS Overseas Research Fellowships and the Department of Physics at MIT 
and Leading Initiative for Excellent Young Researchers, MEXT, Japan.
M.~Y. was also supported by the U.S. Department of Energy, Office of Science, Office of High Energy Physics of U.S. Department of Energy under grant Contract Number DE-SC0012567.
M.~Y. thanks the hospitality during his stay at DESY. 
\end{acknowledgments}

\bibliography{references}

\end{document}